\documentclass[twocolumn,preprintnumbers,amssymb,amsmath,superscriptaddress,letterpaper,nofootinbib]{revtex4}[12pt]
\usepackage{times,graphics}
\usepackage[colorlinks]{hyperref}

\usepackage{graphicx}
\usepackage{dcolumn}
\usepackage{bm}
\usepackage{natbib}
\usepackage{pstricks}
\usepackage{epsfig}
\usepackage{epstopdf}
\usepackage{multirow}
\usepackage{amsmath}
\usepackage{lipsum}

\newcommand{\be}{\begin{equation}}
\newcommand{\ee}{\end{equation}}
\newcommand{\bea}{\begin{eqnarray}}
\newcommand{\eea}{\end{eqnarray}}

\makeatletter
\newcommand*{\shifttext}[2]{%
	\settowidth{\@tempdima}{#2}%
	\makebox[\@tempdima]{\hspace*{#1}#2}%
}
\makeatother

\begin{document}

%\title{ $N$ (randomly) coupled (massive) gravitons:\\  $L_{EH}=M_p^2\sqrt{g}\,R$ and $\Lambda$ emerge naturally from electroweak energy scale}
%\title{NGravity: $N$ (randomly) coupled (massive) gravitons\\A resolution on hierarchy problems by theoretically prediction: $\Lambda\,M_p^2\sim M_{EW}^4$}
%\title{a massless graviton emerges naturally from infinite randomly coupled gravitons}
%\title{NGravity: $N$ (randomly) coupled (massive) gravitons\\more than just reducing two hierarchy problems to one and even less}
%\title{NGravity: $N$ (randomly) coupled (massive) gravitons\\emergence of Einstein-Hilbert action and a resolution for hierarchy problems}
\title{Large-$N$ Random Matrix Gravity and the Double Hierarchy Problem 
}

\author{Nima Khosravi}
\email{n-khosravi@sbu.ac.ir}
\affiliation{Department of Physics, Shahid Beheshti University, G.C., Evin, Tehran 19839, Iran}

\date{\today}

\begin{abstract}
Why are the cosmological constant, electroweak and Planck scales so different? This ``double hierarchy" problem, where $\Lambda \ll M^2_{EW} \ll M^2_p$, is one of the most pressing in fundamental physics.  We show that in a theory of $N$ randomly coupled massive gravitons at the electroweak scale, these scales are linked precisely by such a double hierarchy for large $N$, with intriguing cosmological consequences.  Surprisingly, in all the physical scales, only one massless graviton emerges which is also, effectively, the only one that is coupled to matter, giving rise to standard Einstein gravity, with $M_p^2\, G_{\mu\nu}= T_{\mu\nu}$ at large $N$.   In addition there is a tower of massive gravitons, the lightest of which can drive late-time acceleration. In this scenario, the observed empirical relation $\Lambda\, M_p^2 \sim M_{EW}^4$ as well as the double hierarchy, arise naturally since $\Lambda \sim M^2_{EW}/\sqrt{N}$ and $M^2_p \sim \sqrt{N}M_{EW}^2$.

\end{abstract}
\maketitle

\section{Introduction:}
In this work, a model is proposed to connect two main hierarchy problems in theoretical physics, the electroweak and the cosmological constant, by using Random Matrix Theory for the gravitational sector. The Einstein-Hilbert action naturally emerges with the coupling to matter at Planck scale, $M_p^2$. In addition, a tower of massive graviton  is predicted which can also be a resolution for the hierarchy problems.  This tower of massive gravitons, due to their negligible interaction with visible matter, can describe the dark sector of our universe with their own cosmological footprints. We emphasize that our idea could be applied to other fields of physics but here we focus on the gravitational sector. Now let us review the status of our understanding of the gravitational interaction before presenting the main idea.

The Einstein-Hilbert action describes our gravitational sector very successfully. Theoretically, it predicts a massless spin-2 particle, graviton, as the mediator of the gravitational field. Its coupling to matter is extraordinarily weak, given by the inverse Planck mass squared, $1/M^2_p$. On the other hand, (observable) matter content of the universe is described by the standard model of particle physics at electroweak energy scale, $M^2_{EW}$. The gap between these two energy scales is huge and is one of the famous hierarchy problems formulated in theoretical physics. On the other hand, cosmological observations support a model which is almost dark. Dark matter is needed for structure formation as well as describing the cosmic microwave background and up to now there is no hint for their direct detections. The nature of dark energy, $\Lambda$, which is responsible for the late time accelerating phase of the universe is almost unknown. The existence of a non-zero but very tiny $\Lambda$ causes a new hierarchy between involved energy scales in the universe as $\Lambda\ll M_{EW}^2\ll M_p^2$. It is worth to mention here that there is a very peculiar empirical relation between these energy scales, ``$\Lambda\, M_p^2\sim M_{EW}^4$", without any theoretical justification.

On the other hand there is always the question that whether our universe could be in another form. This question can be asked if it is possible to describe our own universe with its current physics starting from a random framework? This  brings the well-known theory of Random Matrix Theory (RMT) to our mind. Wigner introduced random matrices in nuclear physics \cite{wigner} but the influential works have been done by Dyson \cite{dyson}. The applications of RMT are widespread from quantum chaos \cite{q-chaos} to quantum gravity \cite{q-gravity}. This field was also studied in pure mathematics. For our purpose it is specially useful to mention two seminal theorems in this field. First, the eigenvalue spectrum of an $N\times N$ random matrix has a maximum eigenvalue at ${\cal{O}}(N/2)$ while the other smaller eigenvalues are distributed in Wigner's semi-circle $[-\sqrt{N},\sqrt{N}]$ \cite{random-eigen}. Second, the Perron-Frobenius theorem  states that for an $N\times N$ matrix with positive real components there is a largest eigenvalue whose corresponding eigenvector has all positive components and it is not true for any other eigenvectors.

In a promising work, Sachdev and Ye in \cite{sachdev} studied a network of spins where the couplings are chosen from a random distribution. Kitaev generalized their idea \cite{kitaev}, nowadays well-known  as SYK model, which is related to the black-hole entropy and AdS/CFT correspondence \cite{Maldacena:2016hyu}. In SYK model the interaction Hamiltonian for $N$ Majorana fermions, $\psi_i$'s, is written as $H=\sum_{ijkl}\gamma_{ijkl}\psi_i\psi_j\psi_k\psi_l$ where the couplings $\gamma_{ijkl}$ are taken randomly from a Gaussian distribution. The final result in SYK model has a $1/N$ expansion where the zeroth order term is uniquely determined for large $N$ limit independent of the initially randomly chosen $\gamma_{ijkl}$ \cite{Maldacena:2016hyu}. 

Inspired by the above idea, in this work we will study a multi-massive-graviton model\footnote{It is well-known that gravity cannot be explained by a scalar or a vector field. This means having $N$ spin-2 particle can be interpreted as the most general scenario for describing gravitational field.}, where all the couplings are taken from a random distribution. 
It will be shown that the results are promising: i) we can answer why Einstein-Hilbert action governs the gravitational force in our universe\footnote{This question has been studied in \cite{Khosravi:2016kfb,Khosravi:2017aqq} but by a different viewpoint.}, ii) our model gives a theoretical justification for $\Lambda\, M_p^2\sim M_{EW}^4$ which makes two independent hierarchy problems one, iii) in addition we predict a tower of massive gravitons\footnote{This property of our model, i.e. having large number of new degrees of freedom, is very similar to what has been studied in \cite{Arkani-Hamed:2016rle,Dvali:2009ne,Dvali:2007hz} in other contexts.} between $\Lambda$ and $M_p^2$ which can be interpreted as dark sector since their interaction with visible matter is negligible.

\section{The Multi-Graviton Model:}
The (generalized) quadratic Lagrangian for $N$ massive spin-2 particles can be written as perturbations of the metric around the Minkowski\footnote{Our analysis can be easily generalized to a general background metric $\bar{g}_{\mu\nu}$.} one,  $g_{\mu\nu}^{(i)}=\eta_{\mu\nu}+h_{\mu\nu}^{(i)}$, as 
\begin{eqnarray}\label{Lag-mixed}
{\cal S}&=&\int dx^4\,\bigg[M^2\,\sum_{i=1}^N\sum_{j=1}^N\bigg(\,K_{ij}\,h^{(i)}_{\mu\nu}\, {\cal E}^{\mu\nu\alpha\beta}\,h^{(j)}_{\alpha\beta} \,\bigg)\\\nonumber&+&M^4\,\sum_{i=1}^N\sum_{j=1}^N\,{m}_{ij}\,V(h^{(i)}_{\mu\nu},h^{(j)}_{\alpha\beta})+\sum_{i=1}^{N}\alpha_i\,h^{(i)}_{\mu\nu}T^{\mu\nu}\bigg]
\end{eqnarray}
where ${\cal E}^{\mu\nu\alpha\beta}$ is Lichnerowicz (second order differential) operator. We also assumed mixing kinetic terms (by non vanishing off-diagonal terms of $K_{ij}$) which makes the above Lagrangian more general than the usual one\footnote{We only assume that $K_{ij}$ is positive definite. This assumption guarantees that all the eigenvalues are positive and consequently the kinetic term does not produce any ghosts.}. In the above Lagrangian $K_{ij}$ and ${m}_{ij}$ are ${\cal O}(1)$ dimensionless coefficients and we emphasize that there is just one mass scale, $M^2$, in this action. The last term shows the coupling between gravity and matter sectors and we assume  $\alpha_i=1$ for all $i$'s\footnote{In this work we focus on $\alpha_i=1$ case but in a general scenario $\alpha_i$'s could be ${\cal{O}}(1)$ quantities. This generalization needs assumption on the distribution functions of in this work we focus on $\alpha_i$'s and make the analysis harder and case dependent. It will remains for the future works.}. In the matrix notation we can rewrite the kinetic and mass term as
\begin{eqnarray}\label{kin-mass-non-diagonal}
	M^2\,\mathbf{h^T}\,\mathbb{K}\,{\cal E}\,\mathbf{h}\,+\,M^4\,\mathbf{h^T}\,\mathbb{M}\,\mathbf{h}
\end{eqnarray}
where ${\cal E}$ represnts ${\cal E}^{\mu\nu\alpha\beta}$. $\mathbb{K}$  is the kinetic matrix
\begin{equation}\label{kinetic}
	\mathbb{K}\equiv
	\left(\begin{smallmatrix}
		K_{11} & K_{12} & \hdots   & \hdots    & K_{1N} \\
		K_{21} & \ddots   & \ddots  & \ddots & \vdots \\
		\vdots &    & \ddots & \ddots & \vdots  \\
		\vdots &    &        & \ddots & \vdots \\
		K_{N1} &\hdots   & \hdots  & \hdots &  K_{NN}
	\end{smallmatrix}\right)_{N\times N},
\end{equation}
while the mass matrix $\mathbb{M}$ is shown as
\begin{equation}
	\mathbb{M}\equiv
	\left(\begin{smallmatrix}
		m_{11} & m_{12}   & \hdots & \hdots   & m_{1N} \\
		m_{21} & \ddots   & \ddots  & \ddots & \vdots \\
		\vdots &    & \ddots& \ddots  & \vdots  \\
		\vdots &    &        & \ddots & \vdots \\
		m_{N1}    & \hdots & \hdots & \hdots &  m_{NN}
	\end{smallmatrix}\right)_{N\times N}
\end{equation}
and we have 
\begin{equation}
	\mathbf{h}\equiv
	\left(\begin{smallmatrix}
		h^{(1)}_{\mu\nu}\\
		h^{(2)}_{\mu\nu}   \\
		h^{(3)}_{\mu\nu}  \\
		\\
		\vdots \\
		\\
		\\
		h^{(N)}_{\mu\nu} 
	\end{smallmatrix}\right)_{N\times 1}.
\end{equation}
Note that $K_{ij}$'s and $m_{ij}$'s are dimensionless parameters.

\subsection{Random Kinetic and Mass Matrices:}
We would like to assume that the components of the kinetic and mass matrices are random numbers belong to $[0,1]$. This means that there is no priory assumption for these matrices and all of the coefficients in (\ref{Lag-mixed})  are at order ${\cal O}(1)$, i.e., they are natural\footnote{Note that it is easy to see that the main results are not changed if one define natural parameters e.g. in $[0.1,10]$ instead of $[0,1]$. The only difference is that a factor of $\sim 10$ should be multiplied to eigenvalues. Note that one can also change uniform distribution $[0,1]$ to a Gaussian one and if the mean and variance be ${\cal{O}}(1)$ the  results are not modified for large $N$'s which is our goal in this work.}. For the first step let's focus on the kinetic term (\ref{kinetic}) and try to make it diagonal. F\"uredi-Koml\`os theorem emphasizes that for an $N\times N$ matrix with random components in $[0,1]$ there is just one very large eigenvalue, $\lambda_0 \sim \frac{{\cal O}(N)}{2}$, when the other eigenvalues, $\lambda_{i\neq0}$'s, are distributed around zero on Wigner semi-circle with radius ${\cal O}(\sqrt{N})$. So for large $N$'s we have $|\lambda_{i\neq0}|\ll \lambda_0$. The key property of the largest eigenvalue, $\lambda_0$, is that the components of its eigenvector are all positive and it is not true for other $\lambda_i$'s eigenvectors according to Perron-Frobenius theorem. By diagonalizing the kinetic matrix using the matrix of eigenvectors we get
\begin{eqnarray}\label{Lag-diag}\nonumber
{\cal L}_{kin}&=&M^2\bigg[N\,E_{\mu\nu}\, {\cal E}^{\mu\nu\alpha\beta}\,E_{\alpha\beta} \,+\,\sqrt{N}\,g^{(1)}_{\mu\nu}\, {\cal E}^{\mu\nu\alpha\beta}\,g^{(1)}_{\alpha\beta}\\\nonumber&+&\sum_{i=2}^{N-1}\mu_i(N) g^{(i)}_{\mu\nu}\, {\cal E}^{\mu\nu\alpha\beta}\,g^{(i)}_{\alpha\beta}\bigg]
\end{eqnarray}
where $\mu_i(N)/N$ are decreasing faster than $1/\sqrt{N}$. In the above Lagrangian $E_{\mu\nu}$ corresponds to the largest eigenvalue and $g^{(i)}_{\mu\nu}$'s are related to smaller eigenvalues. For future purposes we picked out $g^{(1)}_{\mu\nu}$ which corresponds to the second largest eigenvalue, i.e., $\sim \sqrt{N}$. As we mentioned previously, the eigenvector's components of the largest eigenvalue are all positive which means $E_{\mu\nu}=\sum_{i=1}^{N}a_ih^{(i)}_{\mu\nu}$ where $a_i>0$. 

If we assume the orthogonal matrix $\mathbf{S}$ is built from the eigenvectors of $\mathbb{K}$ then it gives $\mathbf{S}\,\mathbb{K}\,\mathbf{S^T}$ is diagonal. Now we can rewrite the kinetic part in (\ref{kin-mass-non-diagonal}) as
\begin{eqnarray}
	M^2\,\mathbf{(S\,h)^T}\,\big(\mathbf{S}\,\mathbb{K}\,\mathbf{S^T}\big)\,{\cal E}\,(\mathbf{S\,h})
\end{eqnarray}
which gives a new coordinate ``$\mathbf{S\,h}$" to work with. Note that $E_{\mu\nu}$ and $g_{\mu\nu}^{(i)}$'s are introduced in this new coordinate. In this new basis the mass term can be written as
\begin{eqnarray}
	M^4\,\mathbf{(S\,h)^T}\,\big(\mathbf{S}\,\mathbb{M}\,\mathbf{S^T}\big)\,(\mathbf{S\,h}).
\end{eqnarray}
Since the $\mathbb{M}$'s components, $m_{ij}$'s, are random then we expect to the components of $\mathbf{S}\,\mathbb{M}\,\mathbf{S^T}$ to be random too. In addition since the $\mathbf{S}$ is an orthogonal matrix then the components of mass term in this new coordinate has the same order of magnitude as their original values i.e. ${\cal{O}}(1)$.  This means we can write the kinetic and mass term in new basis as
\begin{eqnarray}\label{final-ideal}\nonumber
	{\cal L}&=&N\,M^2\bigg[\,E_{\mu\nu}\, {\cal E}^{\mu\nu\alpha\beta}\,E_{\alpha\beta} \,+\,\frac{1}{\sqrt{N}}\,g^{(1)}_{\mu\nu}\, {\cal E}^{\mu\nu\alpha\beta}\,g^{(1)}_{\alpha\beta}\\\nonumber&&\hspace{1cm}+\sum_{i=2}^{N-1}\frac{\mu_i(N)}{N} g^{(i)}_{\mu\nu}\, {\cal E}^{\mu\nu\alpha\beta}\,g^{(i)}_{\alpha\beta}\bigg]\\\nonumber
	&+&N\,M^4\bigg[\frac{1}{N}\,E_{\mu\nu}\, {P}^{\mu\nu\alpha\beta}\,E_{\alpha\beta}+\frac{1}{N}\,g^{(1)}_{\mu\nu}\, {P}^{\mu\nu\alpha\beta}\,g^{(1)}_{\alpha\beta} \,\\\nonumber
	&+&\frac{1}{N}\sum_{i=2}^{N-1} g^{(i)}_{\mu\nu}\, {P}^{\mu\nu\alpha\beta}\,g^{(i)}_{\alpha\beta}+\frac{1}{N}\sum_{i\neq j}^{N} g^{(i)}_{\mu\nu}\, {Q}^{\mu\nu\alpha\beta}\,g^{(j)}_{\alpha\beta}\bigg],
\end{eqnarray}
where ${P}^{\mu\nu\alpha\beta}$ and ${Q}^{\mu\nu\alpha\beta}$ are representing mass term and the mixing between the different gravitons respectively (and they are given by the background metric $\eta_{\mu\nu}$).

\subsection{Coupling to Matter}

Now let's rewrite the coupling to matter, i.e. $\sum_{i=1}^{N}\,\alpha_ih^{(i)}_{\mu\nu}T^{\mu\nu}$ in (\ref{Lag-mixed}), in the terms of our new coordinates $E_{\mu\nu}$ and $g^{(i)}_{\mu\nu}$. As we already mentioned we assume $\alpha_i=1$ which means the coupling to matter can be written as  $\big(\sum_{i=1}^{N}\,h^{(i)}_{\mu\nu}\big)T^{\mu\nu}$. Now the question will be how a vector as $\vec{\alpha}=(1,1,...,1)$ can be written as a linear combination of eigenvectors of matrix ${\mathbb{K}}$ in (\ref{Lag-mixed})? We claim that the answer is that the main contribution comes from $E_{\mu\nu}=\sum_{i=1}^{N}a_ih^{(i)}_{\mu\nu}$ and all $g^{(i)}_{\mu\nu}$'s contribution is at $1/\sqrt{N}$ order correction.

Let's multiply matrix ${\mathbb{K}}$ and vector $\vec{\alpha}$ to have $v_i=\sum_{j=1}^{N}K_{ij}\alpha_j$. Consequently, we have $v_i=\sum_{j=1}^{N}K_{ij}$ since $\alpha_i\sim 1$, i.e. each component of $v_i$ is the sum of all the components of $i$'th row of matrix ${\mathbb{K}}$. Now since the components of ${\mathbb{K}}$ are random numbers then the ``law of large numbers" says that all the $v_i$'s should be $N\times\frac{1}{2}+{\cal{O}}(\sqrt{N})$, for very large $N$'s, since $1/2$ is the mean value of our random distribution. Note that the correction is decreasing as $1/\sqrt{N}$ in comparison to the mean value. In matrix form we have
\begin{eqnarray}\label{N-eigen}
	{\mathbb{K}}_{N\times N}\,.\,
	\left(\begin{smallmatrix}
		1\\
		1\\
		. \\
		.\\
		1
	\end{smallmatrix}\right)_{N\times 1}=
\frac{N}{2}\left[\left(\begin{smallmatrix}
	1\\
	1\\
	. \\
	.\\
	1
\end{smallmatrix}\right)_{N\times 1}+
\frac{2}{\sqrt{N}}
\left(\begin{smallmatrix}
	{\cal O}(1)\\
	{\cal O}(1)\\
	. \\
	.\\
	{\cal O}(1)
\end{smallmatrix}\right)_{N\times 1}\right].
\end{eqnarray}
For $N\rightarrow\infty$, we can ignore $1/\sqrt{N}$ term which means $(1,1,...,1)$ is $\mathbb{K}$'s eigenvector with eigenvalue $N/2$. This means for $N\rightarrow\infty$, this eigenvector corresponds exactly to $E_{\mu\nu}$ which is introduced above. So in this limit the coupling to matter will be
\begin{eqnarray}
	\sum_{i=1}^{N}\,h^{(i)}_{\mu\nu}T^{\mu\nu}=\sqrt{N}\,E_{\mu\nu}\,T^{\mu\nu}.
\end{eqnarray}
Now if we assume $N$ is large but not infinity, we expect that the $1/\sqrt{N}$ correction in (\ref{N-eigen}) should be take into account\footnote{In (\ref{N-eigen}) we have to divide both side to $\sqrt{N}$ to make the vectors, unit vectors.}
\begin{eqnarray}
	\sum_{i=1}^{N}\alpha_i\,h^{(i)}_{\mu\nu}T^{\mu\nu}=\sqrt{N}\bigg[E_{\mu\nu}\,T^{\mu\nu}+{\cal O}(1/\sqrt{N})\bigg].
\end{eqnarray}
This means the contribution of other metrics $g^{(i)}_{\mu\nu}$, all together, gives ${\cal O}(1/\sqrt{N})$ correction.

\subsection{Emerged gravity model from large-$N$ coupled gravitons:}
Now let's rewrite the action (\ref{Lag-mixed}) according to our above discussion. By rescaling the metrics, as $g^{(i)}_{\mu\nu}\rightarrow  \texttt{h}^{(i)}_{\mu\nu}/\alpha_i(N)$ where $\alpha^2_i(N)=\mu_i(N)/N$ and $\mu_1(N)=\sqrt{N}$,  we can make all the kinetic terms canonical  
\begin{eqnarray}\label{final-ideal-1}\nonumber
{\cal L}&=&N\,M^2\bigg[\bigg(\,E_{\mu\nu}\, {\cal E}^{\mu\nu\alpha\beta}\,E_{\alpha\beta}+M^2\frac{1}{N}\,E_{\mu\nu}\, {P}^{\mu\nu\alpha\beta}\,E_{\alpha\beta}\bigg)\\\nonumber&+&\bigg(\,\texttt{h}^{(1)}_{\mu\nu}\, {\cal E}^{\mu\nu\alpha\beta}\,\texttt{h}^{(1)}_{\alpha\beta}+M^2\frac{1}{\sqrt{N}}\,\texttt{h}^{(1)}_{\mu\nu}\, {P}^{\mu\nu\alpha\beta}\,\texttt{h}^{(1)}_{\alpha\beta}\bigg)  \\\nonumber&+&\sum_{i=2}^{N-1} \bigg(\texttt{h}^{(i)}_{\mu\nu}\, {\cal E}^{\mu\nu\alpha\beta}\,\texttt{h}^{(i)}_{\alpha\beta}
+M^2\frac{1}{\mu_i(N)} \texttt{h}^{(i)}_{\mu\nu}\, {P}^{\mu\nu\alpha\beta}\,\texttt{h}^{(i)}_{\alpha\beta}\bigg)\\\nonumber
&+&M^2\sum_{i\neq j}^{N} \frac{1}{\sqrt{\mu_i(N)\mu_j(N)}}\, \texttt{h}^{(i)}_{\mu\nu}\, {Q}^{\mu\nu\alpha\beta}\,\texttt{h}^{(j)}_{\alpha\beta}\bigg]\\\nonumber
&+&\sqrt{N}\bigg[E_{\mu\nu}\,T^{\mu\nu}+{\cal O}(1/\sqrt{N})\bigg].
\end{eqnarray}
Note that the correction term in the last bracket says all the other metrics contribution in coupling to matter is at order ${\cal O}(1/\sqrt{N})$. 

\section{Main Results:}
The fascinating physics associated with the above mathematical results can be listed as follow
\begin{itemize}
		\item There is an infinite tower of massive gravitons. The second minimum of the mass spectrum is at $\mu^2=\frac{1}{\sqrt{N}}M^2$ for the metric labeled $\texttt{h}_{\mu\nu}^{(1)}$.
	
	\item The Planck mass is the ratio of the coefficients of kinetic term and coupling term to matter i.e. $M_p^2\equiv \sqrt{N}\,M^2$. 
	
	\item A simple algebra shows that $\boxed{\mu^2\,M_p^2=M^4}$ . This is an amazing result:\\  Up to now we haven't said anything about the \textit{only} mass scale in our model, i.e. $M^2$, but let's fix it to be the electroweak energy scale $M^2=M^2_{EW}$. It is shown in the literature that a massive graviton (with small mass) can cause the late time acceleration with a cosmological constant as $\Lambda\sim \mu^2$ \cite{Khosravi:2011zi,Gumrukcuoglu:2011ew,Comelli:2011zm,Volkov:2011an,vonStrauss:2011mq}. So the above relation becomes $\Lambda\,M_p^2\sim M_{EW}^4$ which relates the electroweak, Planck and cosmological constant energy scales together. 
	This is an interesting result which relates two (independent) hierarchy problems between ``$M^2_p$ and $M^2_{EW}$" and  ``$M^2_p$ and $\Lambda$" to just one problem: why $N$ is large? This means the physics of the largest and the smallest energy scales are connected. Note that these two scales emerge from $M^2_{EW}$ scale naturally. It is worth to mention that $\Lambda\, M_p^2\sim M^4_{EW}$ is supported by observations although there was not too much theoretical justification for it\footnote{There is an interesting idea by C. Burgess, supersymmetric large extra dimension (SELD) \cite{Burgess:2004ib,Burgess:2004kd}, where the $M_p$ and $M_{EW}$ become related as $M_{EW}^2\sim M_p / r$ where $r$ is circumference of two  extra dimensions.}.

	\item Now by setting $\Lambda\sim M^2/\sqrt{N}$ the mass for $E_{\mu\nu}$ will be $m^2_E\sim\Lambda/\sqrt{N}$. This mass is not observable in the physical scales since its corresponding wavelength is much larger than the present Hubble scale given by inverse $\Lambda$. This means, for large $N$'s, the $E_{\mu\nu}$ presents the only  ``\textit{massless}" graviton $E_{\mu\nu}$ in all the observable scales, which is emerged automatically.

	\item The ``\textit{massless}" graviton has the main contribution in coupling to (visible) matter. All the other metrics together give ${\cal O}(1/\sqrt{N})$ correction to it which is very tiny. This is a very interesting result: the emergence of the Einstein-Hilbert gravity is inevitable in random matrix gravity.

    \item In addition, a tower of massive gravitons arise naturally between $[\mu^2,M_p^2]$, This can be seen as a resolution for the hierarchy problem by prediction of new physics between $M^2_{EW}$ and $M_p^2$ scales. 
\end{itemize}

\subsection{Additional Byproducts:}
In addition to the above main results there are some further implications in our model. They are important and need more considerations in details for future but it is worth to be listed:

\begin{itemize}
	\item Since the tower of massive gravitons are very weakly coupled to matter they can be interpreted as dark components of the universe i.e. dark matter and dark energy.  
	\item The massive gravitons with masses close to $\mu^2$ are the ones comparable with the inverse Hubble scale. These massive gravitons cannot be localized and  results in self-accelerating solutions describing the late time acceleration\footnote{Note that there are many massive gravitons with masses around $\mu^2$ but they may modify the late time cosmological constant just by a factor of ${\cal O}(1)$. It is obvious from the results in \cite{Khosravi:2011zi} by assuming $m^2_3=0$.}.  
    \item There are many other massive gravitons in this model with the masses above $\mu^2$ which are larger than the inverse Hubble scale which can be interpreted as dark matter particles.  It should be mentioned that the massive spin-2 particles as dark matter are suggested and studied in \cite{Babichev:2016bxi,Babichev:2016hir,Albornoz:2017yup,Aoki:2016zgp}. In these works under some assumptions the stability of dark matter particles has been considered. It is concluded that massive spin-2 particles can be a candidate for stable cold dark matter particles. These spin-2 particles can have very small masses while they are not relativistic necessarily, since they do not contribute in thermal bath. This means they are not constrained by CMB bounds\footnote{The well-known similar scenario is axions, particles that can have mass even less than $meV$ energy scale but since they do not have interaction with visible matter, they do not contribute in thermal bath. This makes them non-relativistic even if their masses are very small \cite{Marsh:2015xka,Duffy:2009ig}.}.
\end{itemize}

\section{Concluding Remarks and Future Perspectives:}
To recapitulate, it has been shown that from $N$ randomly coupled massive gravitons, at $M^2_{EW}$ energy scale, the Einstein-Hilbert action (with an effectively massless graviton in all physical scales) emerges automatically. This massless graviton is coupled to matter at $M_p^2$ energy scale while the other metrics give a  correction tp to  ${\cal O}(1/\sqrt{N})$ which is very negligible for large $N$. This means in our Random Matrix Gravity theory, the emergence of Einstein-Hilbert gravity is unavoidable.  More interestingly, the empirical relation $\Lambda\,M_p^2\sim M_{EW}^4$ finds a theoretical justification in our model. This result amazingly reduces two fundamental hierarchy problems to just one. On the other hand a tower of massive gravitons also emerges naturally such that  $m^2\in [\Lambda,M^2_p]$ which can be a solution for hierarchy problem.  In addition the tower of massive gravitons can be responsible for dark sector of our universe due to their very weakly coupling to visible matter.

We believe that Random Matrix Gravity can be a way to think about the hierarchy problems and its  outstanding results makes it an interesting model to consider more. One natural theoretical direction of pursuing is embedding this model into the non-linear multi-graviton models\footnote{It has been shown that in multi-gravity models, there is always one exact massless graviton (see e.g. \cite{Khosravi:2011zi}). We claim that this massless graviton is our $E_{\mu\nu}$ up to ${\cal{O}}(1/\sqrt{N})$ corrections. However it needs further considerations in the future. } \cite{Hinterbichler:2012cn,Noller:2013yja,Scargill:2014wya,Alberte:2019lnd} which is based on ghost free massive gravity \cite{deRham:2010kj,Hassan:2011hr}.This is crucial to check under which conditions our model avoid ghosts. For this purpose the vierbein formalism seems more suitable since the dRGT mass term's structure is more transparent. Another direction of research is studying our model in the framework of clockwork idea. For this purpose, \cite{Niedermann:2018lhx} may be useful where the clockwork idea \cite{Kaplan:2015fuy,Choi:2015fiu} has been employed for spin-2 particles and its non-linear extension is also studied\footnote{We thank Johannes Noller for mentioning these works to us and also the similarity between our model and the clockwork idea.}. As it is also stated in \cite{Niedermann:2018lhx}, there is no radiative instability caused by coupling to matter since there is only one metric which is coupled to matter. However the loops in graviton interactions with each others should be studied which remain for the future works.  On the other hand the phenomenology of a tower of spin-2 particles is very rich in cosmology and the physics of dark matter \cite{Babichev:2016bxi,Babichev:2016hir,Albornoz:2017yup,Aoki:2016zgp} which is also predicted in \cite{Niedermann:2018lhx}. The observational smoking gun of our model can be pursued in the interaction between the gravitons which may results in interacting dark matter as well as dynamical dark energy. In addition there is a very weak coupling between our massive gravitons (i.e. dark sector) and the visible matter which can be another fingerprint of our model.

\section*{Acknowledgments}
I would like to thank Nima Doroud for his comment on \cite{Khosravi:2016kfb} and its possible relation to \cite{Maldacena:2016hyu} which became the first step of this work. I am grateful to Kasra Alishahi, Shant Baghram,  Marzieh Farhang and Shahab Shahidi for fruitful discussions and commenting on the early drafts. I am grateful to Niayesh Afshordi, Bruce Bassett, Martin Kunz, Johannes Noller and Shahin Sheikh-Jabbari for their comments on the draft. Also I thank Mohsen Amini (Nazanin's father) and Amir Kargaran for their comments on RMT properties. I specially thank Sara Khatibi for her comments on different stages of this work and very fruitful discussions. I am also grateful to the Abdus Salam International Center of Theoretical Physics (ICTP) for their very kind hospitality during the initial steps of this work. I also thank the anonymous referess for her/his comments on the paper.

\end{document}